# Application of symbolic programming for atomic many-body theory

Rytis Juršėnas and Gintaras Merkelis

*Institute of Theoretical Physics and Astronomy of Vilnius University,*
*A. Goštauto 12, LT-01108, Vilnius, Lithuania*
*E-mail address: jursenas@itpa.lt*

**ABSTRACT**

In the present paper by using *Mathematica* system a symbolic programming method for generation of the expressions of the expansion terms of atomic stationary perturbation theory (PT) is presented. For this purpose, the package named as *NCoperators* was developed. In producing the PT terms, this package accommodates the features of the Rayleigh-Schrödinger perturbation theory (RSPT), the second quantization method (SQR) and angular momentum theory (AMT). The package was built in such a way that it could be applicable in various areas of theoretical atomic spectroscopy. Many functions of the *NCoperators* can be easily adapted for users to their own demand by easily transforming the functions developed. The package gives a possibility of generating formulas in traditional output form. For some quantities the expressions obtained in Mathematica can be converted into *C* code. Although the *NCoperators* package was developed for Unix OS, it can be easily adopted for other OSs if some of the parameters to be changed.

## 1. INTRODUCTION

Atomic Many-Body Perturbation Theory (MBPT), Coupled-Cluster (CC) approach and other theoretical methods provided under the variational or configuration interaction (CI) methodologies, are very well developed and widely explored in many atomic and molecular spectroscopy areas [1] for the study of electron correlation effects [2, 3]. The Rayleigh-Schrödinger PT is the most extensive method used for open-shell atomic structure calculations. A brief review of the latter theory can be found, for example, in [4, 5]. In the above mentioned methods, despite the fact that various symmetries of the system studied are taken into account, the calculation difficulties grow up drastically when one seeks to include electrons correlation effects to achieve the data of high accuracy [6].

For open-shell atoms the unperturbed energy levels are quasidegenerate, therefore the application of the Rayleigh-Schrödinger PT is complicated due to the slow convergence of the PT series. In this case the method of the effective operator developed in [3, 4] is very useful. Here the quasidegenerate states are considered on equal footing by forming the so-called model space while the remaining states (energetically separated from the states of the model space) are treated as a perturbation. Such a separation accelerates the convergence of the PT series. The approach of the effective operator is explored in our study. Note, however, the structure of the operators arising in the PT of the effective operator is more complicated in comparison with the operator of the atomic Hamiltonian used in the RSPT. Nevertheless, the modern calculation techniques of theoretical atomic spectroscopy allow one to construct and calculate the matrix elements of those operators for the open-shell atoms efficiently.

In the recent studies the operators of atomic interactions are presented in the second quantized form. It is well known that the electrons creation and annihilation operators can be interpreted as the spherical tensor operators [7]. This fact gives a possibility to express the operators representing the physical interactions in a coupled tensor form, *i.e.*, in terms of the irreducible tensor products of the creation and annihilation operators. Then the calculation of the matrix element can be done in an efficient way [8, 9]. A coupled tensor product can be obtained for the PT expansion terms of the effective operator too. However, there are a lot of ways to construct a coupled tensor form for the two-, three-, etc. electron operators. In [10] the coupling schemes of the momenta when forming the coupled tensor products which lead to the simplifications of the calculation of the matrix elements have been presented. Generally, the angular decomposition (the summation of the PT expansion terms over the one-electron magnetic quantum numbers) can be realized by using the well known graphical techniques of angular momentum theory (see, for instance, [11]) or algebraically exploiting the nowadays computer programs, such as described, for example, in [12]. The latter approach is developed in the symbolic programming package, called *NCoperators* and presented in this manuscript. The package realizes the systematic application of the methodology for the angular decomposition of the PT series and the coupling schemes of the ranks for the operators given in [9, 10].

## 2. MATHEMATICAL BACKGROUND

In this section we briefly describe the method used by us to investigate the electron correlation effects for the energy spectra of the open-shell atoms. The comprehensive description of this methodology can be found in [4]. The implementation of this approach programmed in the *NCoperators* package is given in the next section (RSPT block).
We start from the atomic Hamiltonian $H$ which reads

$$H = H_0 + V, \tag{1}$$

where

$$H_0 = T + U = \sum_i t_i + \sum_i u_i \tag{2}$$

defines the zero-order approximation. $T$ is the operator of the kinetic energy and the energy of the electron interaction with nucleus. $U$ represents the spherically symmetric potential. The eigenvalue of the operator $t_i + u_i$ is denoted as $\varepsilon_i$. The perturbation operator is expressed as follows

$$V = C - U + H_h = \frac{1}{2} \sum_{i \neq j} g_{ij} - \sum_i u_i + \sum_i h_i. \tag{3}$$

In (3) $C$ represents the two-electron interaction in atoms. It can be the Coulomb or Breit interaction. $H_h$ is given by the one-particle operator and can describe some external perturbation (for example, the magnetic or electric field) or the hyperfine interaction.
According to [4] the eigenvalues $E$ of the Hamiltonian $H$ can be obtained by solving the eigenvalue equation

$$H_{eff} \phi = E \phi \tag{4}$$

with the eigenfunctions $\phi$. The effective operator $H_{eff} = PV\Omega P$ acts in the space of the model functions $\phi = P\psi$, where the $\psi's$ are the eigenfunctions of the eigenvalue equation $H\psi = E\psi$. $P$ is a projection operator of the so-called model space **P**. The wave operator $\Omega$

$$\psi = \Omega \phi = \sum_{n=0}^{\infty} \Omega^{(n)} \phi, \Omega^{(0)} = 1 \tag{5}$$

is found from the generalized Bloch equation [4]

$$[\Omega, H_0]P = QV\Omega P - \chi PV\Omega P, \chi = \Omega - 1 \tag{6}$$

where $[.,.]$ is a commutator. The $Q = 1 - P$ operator projects the $\psi's$ onto the complementary space **Q** in such a way that $PQ = QP = 0$. Hence, the spaces **P** and **Q** construct the full basis space **I**. The operators $\Omega^{(n)}$ in (5) generate the $n$th order correlation corrections and $\chi$ denotes the correlation operator. For the determination of the PT expansion terms the expressions are written in a second quantized form [4]. Then the quantities under the consideration (for instance, $\Omega^{(n)}, H_{eff}^{(n)}$) are obtained as the products of the one- and two-particle operators expressed in the form

$$F = \sum_{\alpha\beta} a_\alpha a_\beta^\dagger \langle \alpha | f | \beta \rangle, G = \frac{1}{2} \sum_{\alpha\beta\mu\nu} a_\alpha a_\beta a_\nu^\dagger a_\mu^\dagger \langle \alpha\beta | g_{12} | \mu\nu \rangle, \tag{7}$$

where $f$ and $g_{12}$ are the one- and two-particle interaction operators, respectively. The summations run over the quantum numbers $\xi = \alpha, \beta, \mu, \nu$, where $\xi \equiv n_\xi \lambda_\xi m_\xi$ and $\lambda_\xi \equiv l_\xi s, s = \frac{1}{2}$ ($l$ is the orbital quantum number; $m$ is a projection) for *LS*-coupling or $\lambda_\xi \equiv l_\xi j_\xi, j_\xi = l_\xi \pm \frac{1}{2}$ for *jj*-coupling. The quantities $a_\xi, a_\xi^\dagger$ are the electron creation and annihilation operators of the state $\xi$.
According to [4], the one-electron states $\xi$ are decomposed into the three types: core, valence and excited. By saying the core state, usually corresponding to the electrons in the closed subshells, we lay account on the occupied states, while the excited or virtual states are of those absent states in the model space **P**. The valence states which usually form the electron states in the open subshells of the atom, are partially occupied.
The spaces **P** and **Q** are defined in the way so that the operators $O$ appearing in the PT series and expressed as the tensor products of the electrons creation and annihilation operators give the non-zero contributions if the next conditions take place:

i. *POP*: The one-electron states of creation and annihilation operators are of valence type.
ii. *POQ*: 1) The one-electron states of creation operator are of core type; the one-electron states of annihilation operator are of excited or valence type. 2) The one-electron states of creation operator are of core or valence type; the one-electron states of annihilation operator are of excited type.
iii. *QOQ*: The one-electron states of creation and annihilation operators are of all possible types.

Notice, the given conditions also take place and for the operators $PO'P, QO'P, QO'Q$, where $O'=O^\dagger$.

When written in the SQR the perturbation $V$ (see (3)) is expressed as follows
$$V = V_0 + V_1 + V_2, \qquad (8)$$
where the zero- ($V_0$), one- ($V_1$) and two- ($V_2$) body terms are obtained when the Wick's theorem is exploited [4]. According to (6) and the above given definitions of the **P** and **Q** spaces (i-iii), we write for $n>1$
$$H_{eff}^{(n)} = P(V_1 + V_2)\Omega^{(n-1)} P. \qquad (9)$$
In a particular case (see the RSPT block)
$$H_{eff}^{(2)} = P(V_1 + V_2)\Omega^{(1)} P, \qquad [H_0, \Omega^{(1)}]P = Q(V_1 + V_2)P. \qquad (10)$$
When the terms of the PT series are obtained, the angular decomposition is performed. Using the theorems of the angular momentum theory and the technique of the irreducible tensor operators, the expansion terms of the effective operator are constructed over the irreducible coupled tensor products and the *3nj*-coefficients (for the illustration of the construction of the second-order effective operator terms see the SQR and AMT blocks).

## 3. METHODOLOGY

The package *NCoperators* consists of the following main blocks:

1. Rayleigh-Schrödinger Perturbation Theory (RSPT)
2. Second Quantization Representation (SQR)
3. Angular Momentum Theory (AMT)
4. Unstructured External Programming (UEP)

To take all advantage of the SQR and RSPT blocks, one has to compile a free ware package *NCAlgebra* [13]. In the latter package the capacities of multiplications with the non-commutative operators are more flexible than with the standard functions given by *Mathematica*.

### 3.1 RSPT block

The algorithm of the construction of the RSPT block is the following. Firstly, the main functions (operators) used in the atomic many-body perturbation theory (MBPT) are defined. The contractions which appear due to the application of Wick's theorem are studied by the functions named as `OneContraction[]`, `TwoContractions[]`, etc. The Hamiltonian (1), the interaction operator (8), and *n*-order *m*-body wave operator (6) in *NCoperators* are defined as `ZeroOrderHamiltonian` (or alias `H0`), `Potential[n]` ($n=0,1,2$) and `WaveOp[n,m]` (so far $n=1,2$ and $m=1,2,3,4$). The projection operators $P$ and $Q$ are also defined as the quantities `P` and `Q` with the properties studied in the previous section. `BlochEquation[]` (6) is another important function of the RSPT block. There are also the functions which define the effective Hamiltonian (`Heff[]`), normal ordering (`NormalOrder[]`, `NormalOrdering[]`), the operators O acting on different spaces (POP, POQ, QOP, QOQ) with the corresponding rules (`Elimination`) which look after the non-zero terms, energy denominator ($\Delta$`[]`) and many more.

| **Algorithm 1:** Constructing the first-order wave operator $\Omega^{(1)}$ |
|---|
| `NormalOrdering[ SQR[ BlochEquation[ WaveOp[1,""] ] ] ]//. Elimination //. denominator1 //. denominator2` |

When the defined functions are ready to be exploited, the first-order wave operator $\Omega^{(1)}$ (see (10)) is generated. The block-scheme of the construction of $\Omega^{(1)}$ is showed in Algorithm 1. The function `SQR[]` gives the second quantized form of the operators (see (1), (3)). The latter will be discussed in a more detail in the next section.

The functions `denominator1` and `denominator2` produce the energy denominators of the studied terms. The equivalent terms (by means of the equality of their matrix elements) are captured under the functions `sym2H[]`, `sym2V[]` and `sym2VH[]` which correspond to the horizontal, vertical and the composition of the vertical and horizontal mirror symmetries if the terms under consideration are presented by the Goldstone diagrams. Then the rule `style` is applied and the operator is ready for the printing. The $\Omega^{(1)}$ consists of 12 terms (3 one-body and 9 two-body terms).

```
In[2]:= StyledExpressions[WaveOp[1, 1]]
        StyledExpressions[WaveOp[1, 2]]
```

$$\text{Out[2]}= \frac{a_{m_1} a^\dagger_{a_2} \langle m_1 | v | a_2 \rangle}{\varepsilon_{a_2} - \varepsilon_{m_1}} + \frac{a_{r_1} a^\dagger_{a_2} \langle r_1 | v | a_2 \rangle}{\varepsilon_{a_2} - \varepsilon_{r_1}} + \frac{a_{r_1} a^\dagger_{m_2} \langle r_1 | v | m_2 \rangle}{\varepsilon_{m_2} - \varepsilon_{r_1}}$$

$$\text{Out[3]}= -\frac{a_{m_1} a_{n_1} a^\dagger_{a_2} a^\dagger_{b_2} \langle m_1 n_1 | g_{12} | a_2 b_2 \rangle}{2(\varepsilon_{a_2}+\varepsilon_{b_2}-\varepsilon_{m_1}-\varepsilon_{n_1})} - \frac{a_{m_1} a_{n_1} a^\dagger_{a_2} a^\dagger_{n_2} \langle m_1 n_1 | g_{12} | a_2 n_2 \rangle}{\varepsilon_{a_2}-\varepsilon_{m_1}-\varepsilon_{n_1}+\varepsilon_{n_2}} + \frac{a_{n_1} a_{r_1} a^\dagger_{a_2} a^\dagger_{b_2} \langle r_1 n_1 | g_{12} | a_2 b_2 \rangle}{\varepsilon_{a_2}+\varepsilon_{b_2}-\varepsilon_{n_1}-\varepsilon_{r_1}} -$$

$$\frac{a_{n_1} a_{r_1} a^\dagger_{b_2} a^\dagger_{m_2} \langle r_1 n_1 | g_{12} | m_2 b_2 \rangle}{\varepsilon_{b_2}+\varepsilon_{m_2}-\varepsilon_{n_1}-\varepsilon_{r_1}} + \frac{a_{n_1} a_{r_1} a^\dagger_{a_2} a^\dagger_{n_2} \langle r_1 n_1 | g_{12} | a_2 n_2 \rangle}{\varepsilon_{a_2}-\varepsilon_{n_1}+\varepsilon_{n_2}-\varepsilon_{r_1}} + \frac{a_{n_1} a_{r_1} a^\dagger_{m_2} a^\dagger_{n_2} \langle r_1 n_1 | g_{12} | m_2 n_2 \rangle}{\varepsilon_{m_2}-\varepsilon_{n_1}+\varepsilon_{n_2}-\varepsilon_{r_1}} -$$

$$\frac{a_{r_1} a_{s_1} a^\dagger_{a_2} a^\dagger_{b_2} \langle r_1 s_1 | g_{12} | a_2 b_2 \rangle}{2(\varepsilon_{a_2}+\varepsilon_{b_2}-\varepsilon_{r_1}-\varepsilon_{s_1})} - \frac{a_{r_1} a_{s_1} a^\dagger_{a_2} a^\dagger_{n_2} \langle r_1 s_1 | g_{12} | a_2 n_2 \rangle}{\varepsilon_{a_2}+\varepsilon_{n_2}-\varepsilon_{r_1}-\varepsilon_{s_1}} - \frac{a_{r_1} a_{s_1} a^\dagger_{m_2} a^\dagger_{n_2} \langle r_1 s_1 | g_{12} | m_2 n_2 \rangle}{2(\varepsilon_{m_2}+\varepsilon_{n_2}-\varepsilon_{r_1}-\varepsilon_{s_1})}$$

**Figure 1.** The terms of the first-order wave operator

In Figure 1 the indices representing quantum numbers of the one-electron state ξ (see Section 2) are decomposed into core (*a,b,c,d,e,f*), valence (*m,n,p,q,k,l*) and excited (*r,s,t,u,w,x*) states. Note, the summations over the quantum numbers ξ of the observed terms must be fulfilled. For simplicity here and elsewhere the sums are not written.

Further, using (10) the operator $H^{(2)}_{eff}$ can be constructed. Since the wave operator $\Omega^{(1)}$ acts on $Q$ on the left hand side and on $P$ on the right hand side, the interaction operator $V = V_1 + V_2$, where the number of $V_1$ terms is 9 and the number of $V_2$ terms is 81. The algorithm of the construction of $H^{(2)}_{eff}$ is a more complicated task. The operator $PV\Omega^{(1)}P$ is decomposed into the sum of the operators $PV_i\Omega^{(1)}_j P$, where $i,j=1,2$. Then the Wick's theorem is exploited. This is the most time consuming process. After that, the output is partitioned into one-, two- and three-body terms. The example of the resultant output is showed in Figure 2. The rest of the terms can be found in Appendix.

```
In[18]:= H2effectivePrint[[1]]
         H2effectivePrint[[25]]
         H2effectivePrint[[27]]
```

$$\text{Out[18]}= -\frac{a_{p_1} a^\dagger_{m_2} \langle a_1 | v_1 | m_2 \rangle \langle p_1 | v_2 | a_1 \rangle}{\varepsilon_{a_1} - \varepsilon_{p_1}}$$

$$\text{Out[19]}= \frac{a_{n_1} a_{q_1} a^\dagger_{m_2} a^\dagger_{q_2} \langle a_1 n_1 | g_{12} | m_2 p_1 \rangle \langle p_1 q_1 | g_{34} | a_1 q_2 \rangle}{2(\varepsilon_{a_1} - \varepsilon_{p_1} - \varepsilon_{q_1} + \varepsilon_{q_2})}$$

$$\text{Out[20]}= -\frac{a_{n_1} a_{p_1} a_{q_1} a^\dagger_{m_2} a^\dagger_{n_2} a^\dagger_{q_2} \langle a_1 n_1 | g_{12} | m_2 n_2 \rangle \langle p_1 q_1 | g_{34} | a_1 q_2 \rangle}{2(\varepsilon_{a_1} - \varepsilon_{p_1} - \varepsilon_{q_1} + \varepsilon_{q_2})}$$

The one-particle interaction $f$ in (7) is denoted as $v_1$ (for the interaction operator $V_1$) and $v_2$ (for the wave operator $\Omega^{(1)}_1$), whereas the two-particle interaction is denoted as $g_{12}$ (for $V_2$) and $g_{34}$ (for $\Omega^{(1)}_2$). One must keep in mind that the one-particle interaction $f$ is decomposed into two parts. One part is associated with some external fields while the other one represents a two-particle interaction with one contraction.

The definition of the function `Elimination` (Algorithm 1) is formulated according to the conditions (i-iii) given in Section 2.

**Figure 2.** The terms of the second-order effective operator.

Constructing the matrix elements of the generated terms of the second-order effective operator, the creation and annihilation operators are presented as the spherical tensor operators. Such a representation lets one to apply the angular momentum theory and to prepare the matrix element for the calculation in an efficient way when the tensor product of the creation and annihilation operators is written in the irreducible coupled tensor form. The SQR and AMT techniques are also programmed in the *NCoperators* package.

### 3.2 SQR and AMT blocks

The SQR and AMT blocks are closely related and, therefore, they will be considered together. The studied blocks are made of a lot of functions. The creation $a_\xi \equiv a_{m_\xi}^{(\lambda_\xi)}$, annihilation $a_\xi^\dagger \equiv a_{m_\xi}^{(\lambda_\xi)\dagger}$ and transposed annihilation $\tilde{a}_\xi \equiv \tilde{a}_{m_\xi}^{(\lambda_\xi)} = (-1)^{\lambda_\xi - m_\xi} a_{-m_\xi}^{(\lambda_\xi)\dagger}$ operators are presented by `CreatOp[]`, `AnnihOp[]`, `TCreatOp[]`, `TAnnihOP[]`, `ITAnnihOp[]`. These operators satisfy the anticommutation rules which are realized by the functions `AntiCommutator[]` or `TantiCommutator[]`. The example is given in Figure 3. The double star notation marks the non-commutative multiplication.

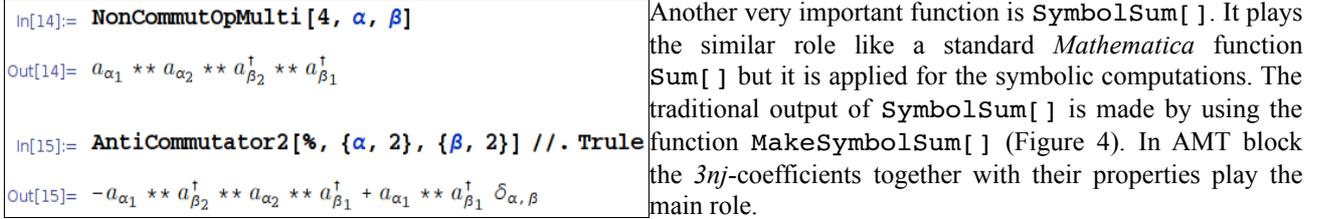

**Figure 3.** The anticommutation rules of the creation and annihilation operators.

Another very important function is `SymbolSum[]`. It plays the similar role like a standard *Mathematica* function `Sum[]` but it is applied for the symbolic computations. The traditional output of `SymbolSum[]` is made by using the function `MakeSymbolSum[]` (Figure 4). In AMT block the *3nj*-coefficients together with their properties play the main role.

The Clebsch-Gordan coefficient is defined as `CGcoeff[]`; *6j* and *9j*-coefficients are defined as `SixJ[]` and `NineJ[]` respectively. Angular momenta recoupling is realized with the functions `Make[]` and `Recoupling[]`.

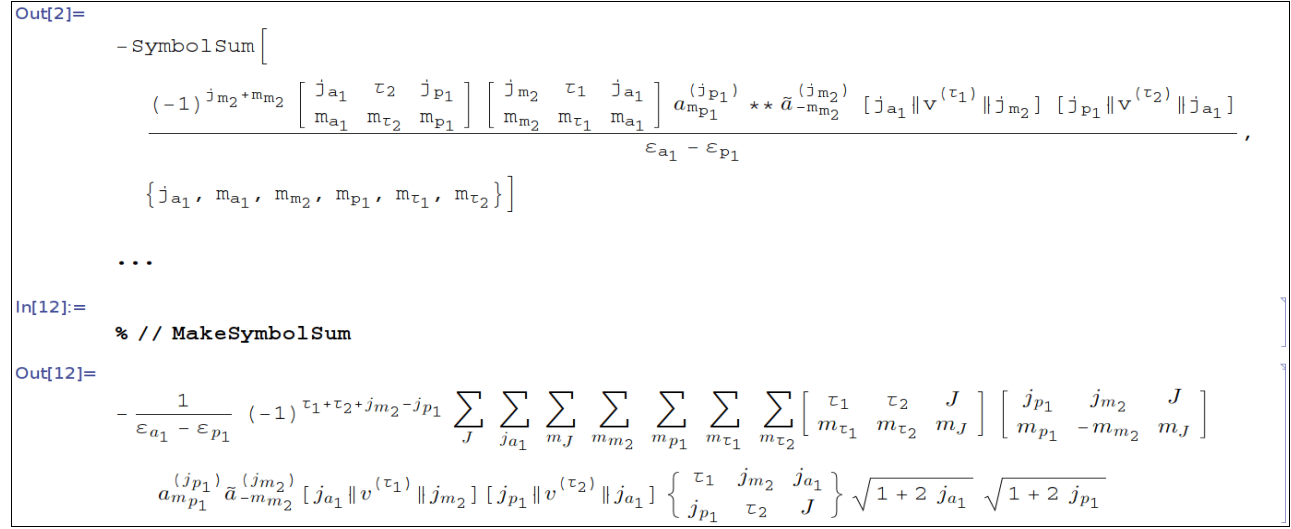

**Figure 4.** The angular decomposition of `Out[18]` given in Figure 2.

The example of the application of the present blocks (SQR and AMT) is showed in Figure 4. The output `Out[2]` is observed when applying the Wigner-Eckart theorem (`WignerEckart[]`), where the quantities [...||...||...] are the reduced matrix elements. The ranks $\tau_1, \tau_2$ correspond to the ranks of $v_1$, $v_2$ in `Out[18]` in Figure 2. Then the momenta recoupling functions `Make[]`, `Recoupling[]` are exploited. In the result (`Out[12]` in Figure 4) the *6j*-coefficient (the quantity with the braces {...}) and the irreducible coupled tensor product $[a^{(j_{p_1})} \times \tilde{a}^{(j_{m_2})}]_{m_J}^{(J)}$ when summing over the projections $m_{p_1}, m_{m_2}$ appear. The traditional output is made by the rule `style`. This becomes very useful when a number of formulas need to be printed. Of course, there is always a possibility to save the output in the LaTeX format which understands the styled output and does not require any additional interventions.

To simplify algebraic manipulations the functions `SumSimplify`, `SumCollect`, `SumExtract`, `SumExpand`,

`KDrule`, `Trule` are used. The rules of the summation of the *3nj*-coefficients over the common momenta are realized by the functions `JRuleOneSixJ[]`, `JRuleOneNineJ[]`, `JRuleTwoSixJ[]`, `JRuleNineJSixJ[]`, `JRuleTwoNineJ[]`, `JruleThreeSixJ[]`. There are also many more functions of specific kind, such as, for example, `Matrixb[]`, `Matrixz[]` (when the two-particle matrix element is expressed in terms of the *b* and *z* coefficients), which insist on a deeper analysis of the mathematical techniques exploited in the atomic spectroscopy. As the main goal of the presented manuscript is to show how the various approaches of the theoretical atomic spectroscopy can be applied under the possibilities given by the computational software system, we shall not discuss them in a detail. However, the analysis of the latter can be found in [8, 10].

### 3.3 UEP block

External programming is a way of communication between *Mathematica* interface and some other (external) programs. Differently from the structured programming which is mainly based on *MathLink*, the unstructured programming does not need any external "tunnels" except for the users own terminal (Unix system).

In our study the external program is *C* compiler (gcc-4.3.3). To make the UEP block of the *NCoperators* package operate, the *Mathematica* header file *mdefs.h* must be put in the directory where all the system headers are placed (in most Unix systems such directory is */usr/include*). This header is supplemented with the functions of calculating *3nj*-coefficients and some other functions (for instance, `ClebschGordanC[]`, `SixJSymbolC[]`, `NineJSymbolC[]`) written in *C* language and necessary in the investigation of the studied quantities. The *Mathematica* should know the definitions of the functions when converting them into the *C* language. For that purpose the extended `cform[ ]` function instead of the standard `CForm[]` is used. Starting from *Mathematica* and employing the UEP block, one can write the headers or even use them for the calculation of the functions considered in the *C* source files. The header files are created by the function `fheader[Cf,f,ftype,var,vartype]`. Here `f` is a *Mathematica* function; `Cf` denotes the corresponding name of the function in the header; `ftype` and `vartype` mark the type of the function and the variables, respectively. These can be *double*, *int*, *char*, etc. `var` is a list of the variables of the function `f`. The idea of the construction of `fheader[]` is very simple. It is presented in Algorithm 2.

| **Algorithm 2:** Writing the header file of the *Mathematica* function |
|---|
| `Block[ {vtv,file,fileh},`<br>`vtv=Cf@@(vartype*var)//TraditionalForm;`<br>`file=ToString[Cf];`<br>`fileh=StringJoin[file,".h"];`<br>`OpenWrite[fileh];`<br>`…`<br>`Close[fileh]; ];` |

Using the standard function `OpenWrite[]` we write the algorithm of the header in the way as it is done in *C* language. The structure of the algorithm depends on, for example, whether the function is of *double* or *void* type. Note that the task when we want to make a *C* source of the given function or to calculate it within *Mathematica* interface, is more complex. Without going into deep discussion of the *C* language properties, here we only notice that Algorithm 2 must be complemented by a simple but very important function `CompileC[]` (or alias `CC[]`) of the UEP block. The algorithm of this function depends on the command by which the compiler is called. In our case it is g++. However, there is no a universal way of creating the algorithm for given functions, because there are a lot of different types of the functions and the corresponding algorithms. Therefore, the UEP block is permanently augmented. A simple usage of the function `Cfunction[]` is showed in Figure 5. The `SixJ[]` and `CGcoeff[]` functions calculating the *6j* and the Clebsch-Gordan coefficients were presented in the AMT block. The comparison of the performing times given by `Timing` function clearly demonstrates that the calculation with the *C* source is faster than with *Mathematica* functions.

```
f[x_, y_, z_, mx_, my_, mz_] :=
    SixJ[{x, y, z}, {x, y, z}]*CGcoeff[{x, mx}, {y, my}, {z, mz}];
f[3., 4., 2., -2., 3., 1.] // N // Timing // Chop
{0.028002, 0.00542328}

Cfunction[f, float, {x, y, z, mx, my, mz}, {3., 4., 2., -2., 3., 1.},
  {float, float, float, float, float, float}] // Timing
{0.016001, 0.0054233}
```

**Figure 5.** The example of the usage of `Cfunction[]`.

Therefore, when solving the main problem with *Mathematica* it is important to have a possibility to delegate some complicated tasks to the *C* source.

## 4. RESULTS AND DISCUSSION

The *NCoperators* package based on *Mathematica* syntax was developed. This package is designed for symbolic computations of various quantities of the theoretical atomic spectroscopy. However, the main purpose of this package is to generate the PT expansion terms and to prepare the obtained expressions, particularly the effective operators, for the calculation of their matrix elements when exploiting the techniques of the second quantization and the angular momentum theories. The angular reduction of the expressions is performed by considering the creation and annihilation operators as the irreducible tensor operators. Subjected on the research area, the package is partitioned into the following blocks: the Rayleigh-Schrödinger perturbation theory (RSPT), the second quantization representation (SQR), the angular momentum theory (AMT) and the unstructured external programming (UEP). The package *NCoperators* was designed in the form that each block can be used separately. The specific feature of the package is that it is based on the algebraic manipulations of the quantities considered in theoretical atomic physics neither the diagrammatic representations developed in the angular momentum theory or the many-body PT. Nevertheless, the present package also gives a possibility to express the terms of the PT series as the Goldstone diagrams up to the second order.

The package has been tested by generating the first- (Figure 1) and the second-order (not presented in this paper) wave operators and the second-order effective Hamiltonian (see the output generated with *NCoperators* in Appendix). Here the symmetry properties of the expansion terms described by the functions `sym2H[]`, `sym2V[]` and `sym2VH[]` of the RSPT block have not been taken into account.

The significant aspect of the developed package is an ability to delegate the tasks which solving with *Mathematica* take the noticeable computer time to the faster computer codes (in our case it is *C* language). Note, the blocks of *NCoperators* are permanently supplemented by the new functions which expand the area of the application of the package and increase the effectiveness of the calculations. Despite operating with a very large number of the expansion terms, the package allows to present them in the standard output which can be simply converted to the LaTeX form.

In conclusion, emphasis that the *NCoperators* package enables the determination of the expressions for the correlation corrections (and calculate them) when investigating various atomic quantities (electron correlation correction for the energy spectra, wave functions, etc.) of the open-shell atoms.

# APPENDIX

The terms of the second-order effective operator without symmetries included generated by the *NCoperators* package.

$$-\frac{a_{p_1} a_{m_2}^\dagger \langle a_1|v_1|m_2\rangle \langle p_1|v_2|a_1\rangle}{\varepsilon_{a_1}-\varepsilon_{p_1}}$$

$$-\frac{a_{n_1} a_{m_2}^\dagger \langle p_1|v_2|a_1\rangle \langle a_1 n_1|g_{12}|m_2 p_1\rangle}{2\left(\varepsilon_{a_1}-\varepsilon_{p_1}\right)}$$

$$\frac{a_{n_1} a_{n_2}^\dagger \langle p_1|v_2|a_1\rangle \langle a_1 n_1|g_{12}|p_1 n_2\rangle}{2\left(\varepsilon_{a_1}-\varepsilon_{p_1}\right)}$$

$$-\frac{a_{n_1} a_{p_1} a_{m_2}^\dagger a_{n_2}^\dagger \langle p_1|v_2|a_1\rangle \langle a_1 n_1|g_{12}|m_2 n_2\rangle}{2\left(\varepsilon_{a_1}-\varepsilon_{p_1}\right)}$$

$$\frac{a_{m_1} a_{m_2}^\dagger \langle p_1|v_2|b_1\rangle \langle m_1 b_1|g_{12}|m_2 p_1\rangle}{2\left(\varepsilon_{b_1}-\varepsilon_{p_1}\right)}$$

$$-\frac{a_{m_1} a_{n_2}^\dagger \langle p_1|v_2|b_1\rangle \langle m_1 b_1|g_{12}|p_1 n_2\rangle}{2\left(\varepsilon_{b_1}-\varepsilon_{p_1}\right)}$$

$$\frac{a_{m_1} a_{p_1} a_{m_2}^\dagger a_{n_2}^\dagger \langle p_1|v_2|b_1\rangle \langle m_1 b_1|g_{12}|m_2 n_2\rangle}{2\left(\varepsilon_{b_1}-\varepsilon_{p_1}\right)}$$

$$\frac{a_{p_1} a_{m_2}^\dagger \langle a_1 b_1|g_{12}|m_2 q_1\rangle \langle p_1 q_1|g_{34}|a_1 b_1\rangle}{4\left(\varepsilon_{a_1}+\varepsilon_{b_1}-\varepsilon_{p_1}-\varepsilon_{q_1}\right)}$$

$$-\frac{a_{p_1} a_{m_2}^\dagger \langle a_1 b_1|g_{12}|m_2 q_1\rangle \langle p_1 q_1|g_{34}|b_1 a_1\rangle}{4\left(\varepsilon_{a_1}+\varepsilon_{b_1}-\varepsilon_{p_1}-\varepsilon_{q_1}\right)}$$

$$-\frac{a_{p_1} a_{n_2}^\dagger \langle a_1 b_1|g_{12}|q_1 n_2\rangle \langle p_1 q_1|g_{34}|a_1 b_1\rangle}{4\left(\varepsilon_{a_1}+\varepsilon_{b_1}-\varepsilon_{p_1}-\varepsilon_{q_1}\right)}$$

$$\frac{a_{p_1} a_{n_2}^\dagger \langle a_1 b_1|g_{12}|q_1 n_2\rangle \langle p_1 q_1|g_{34}|b_1 a_1\rangle}{4\left(\varepsilon_{a_1}+\varepsilon_{b_1}-\varepsilon_{p_1}-\varepsilon_{q_1}\right)}$$

$$-\frac{a_{q_1} a_{m_2}^\dagger \langle a_1 b_1|g_{12}|m_2 p_1\rangle \langle p_1 q_1|g_{34}|a_1 b_1\rangle}{4\left(\varepsilon_{a_1}+\varepsilon_{b_1}-\varepsilon_{p_1}-\varepsilon_{q_1}\right)}$$

$$\frac{a_{q_1} a_{m_2}^\dagger \langle a_1 b_1|g_{12}|m_2 p_1\rangle \langle p_1 q_1|g_{34}|b_1 a_1\rangle}{4\left(\varepsilon_{a_1}+\varepsilon_{b_1}-\varepsilon_{p_1}-\varepsilon_{q_1}\right)}$$

$$\frac{a_{q_1} a_{n_2}^\dagger \langle a_1 b_1|g_{12}|p_1 n_2\rangle \langle p_1 q_1|g_{34}|a_1 b_1\rangle}{4\left(\varepsilon_{a_1}+\varepsilon_{b_1}-\varepsilon_{p_1}-\varepsilon_{q_1}\right)}$$

$$-\frac{a_{q_1} a_{n_2}^\dagger \langle a_1 b_1|g_{12}|p_1 n_2\rangle \langle p_1 q_1|g_{34}|b_1 a_1\rangle}{4\left(\varepsilon_{a_1}+\varepsilon_{b_1}-\varepsilon_{p_1}-\varepsilon_{q_1}\right)}$$

$$-\frac{a_{p_1} a_{q_1} a_{m_2}^\dagger a_{n_2}^\dagger \langle a_1 b_1|g_{12}|m_2 n_2\rangle \langle p_1 q_1|g_{34}|a_1 b_1\rangle}{4\left(\varepsilon_{a_1}+\varepsilon_{b_1}-\varepsilon_{p_1}-\varepsilon_{q_1}\right)}$$

$$\frac{a_{p_1} a_{q_1} a_{m_2}^\dagger a_{n_2}^\dagger \langle a_1 b_1|g_{12}|m_2 n_2\rangle \langle p_1 q_1|g_{34}|b_1 a_1\rangle}{4\left(\varepsilon_{a_1}+\varepsilon_{b_1}-\varepsilon_{p_1}-\varepsilon_{q_1}\right)}$$

$$-\frac{a_{n_1} a_{q_2}^\dagger \langle a_1 n_1|g_{12}|p_1 q_1\rangle \langle p_1 q_1|g_{34}|a_1 q_2\rangle}{2\left(\varepsilon_{a_1}-\varepsilon_{p_1}-\varepsilon_{q_1}+\varepsilon_{q_2}\right)}$$

$$\frac{a_{n_1} a_{q_2}^\dagger \langle a_1 n_1|g_{12}|q_1 p_1\rangle \langle p_1 q_1|g_{34}|a_1 q_2\rangle}{2\left(\varepsilon_{a_1}-\varepsilon_{p_1}-\varepsilon_{q_1}+\varepsilon_{q_2}\right)}$$

$$\frac{a_{p_1} a_{q_1} a_{m_2}^\dagger a_{q_2}^\dagger \langle a_1|v_1|m_2\rangle \langle p_1 q_1|g_{34}|a_1 q_2\rangle}{\varepsilon_{a_1}-\varepsilon_{p_1}-\varepsilon_{q_1}+\varepsilon_{q_2}}$$

$$-\frac{a_{n_1} a_{p_1} a_{m_2}^\dagger a_{q_2}^\dagger \langle a_1 n_1|g_{12}|m_2 q_1\rangle \langle p_1 q_1|g_{34}|a_1 q_2\rangle}{2\left(\varepsilon_{a_1}-\varepsilon_{p_1}-\varepsilon_{q_1}+\varepsilon_{q_2}\right)}$$

$$\frac{a_{n_1} a_{p_1} a_{m_2}^\dagger a_{q_2}^\dagger \langle a_1 n_1|g_{12}|q_1 n_2\rangle \langle p_1 q_1|g_{34}|a_1 q_2\rangle}{2\left(\varepsilon_{a_1}-\varepsilon_{p_1}-\varepsilon_{q_1}+\varepsilon_{q_2}\right)}$$

$$\frac{a_{n_1} a_{q_1} a_{m_2}^\dagger a_{q_2}^\dagger \langle a_1 n_1|g_{12}|m_2 p_1\rangle \langle p_1 q_1|g_{34}|a_1 q_2\rangle}{2\left(\varepsilon_{a_1}-\varepsilon_{p_1}-\varepsilon_{q_1}+\varepsilon_{q_2}\right)}$$

$$-\frac{a_{n_1} a_{q_1} a_{m_2}^\dagger a_{q_2}^\dagger \langle a_1 n_1|g_{12}|p_1 n_2\rangle \langle p_1 q_1|g_{34}|a_1 q_2\rangle}{2\left(\varepsilon_{a_1}-\varepsilon_{p_1}-\varepsilon_{q_1}+\varepsilon_{q_2}\right)}$$

$$-\frac{a_{n_1} a_{p_1} a_{q_1} a_{m_2}^\dagger a_{n_2}^\dagger a_{q_2}^\dagger \langle a_1 n_1|g_{12}|m_2 n_2\rangle \langle p_1 q_1|g_{34}|a_1 q_2\rangle}{2\left(\varepsilon_{a_1}-\varepsilon_{p_1}-\varepsilon_{q_1}+\varepsilon_{q_2}\right)}$$

$$\frac{a_{m_1} a_{q_2}^\dagger \langle m_1 b_1|g_{12}|p_1 q_1\rangle \langle p_1 q_1|g_{34}|b_1 q_2\rangle}{2\left(\varepsilon_{b_1}-\varepsilon_{p_1}-\varepsilon_{q_1}+\varepsilon_{q_2}\right)}$$

$$-\frac{a_{m_1} a_{q_2}^\dagger \langle m_1 b_1|g_{12}|q_1 p_1\rangle \langle p_1 q_1|g_{34}|b_1 q_2\rangle}{2\left(\varepsilon_{b_1}-\varepsilon_{p_1}-\varepsilon_{q_1}+\varepsilon_{q_2}\right)}$$

$$\frac{a_{m_1} a_{p_1} a_{m_2}^\dagger a_{q_2}^\dagger \langle m_1 b_1|g_{12}|m_2 q_1\rangle \langle p_1 q_1|g_{34}|b_1 q_2\rangle}{2\left(\varepsilon_{b_1}-\varepsilon_{p_1}-\varepsilon_{q_1}+\varepsilon_{q_2}\right)}$$

$$-\frac{a_{m_1} a_{p_1} a_{n_2}^\dagger a_{q_2}^\dagger \langle m_1 b_1|g_{12}|q_1 n_2\rangle \langle p_1 q_1|g_{34}|b_1 q_2\rangle}{2\left(\varepsilon_{b_1}-\varepsilon_{p_1}-\varepsilon_{q_1}+\varepsilon_{q_2}\right)}$$

$$-\frac{a_{m_1} a_{q_1} a_{m_2}^\dagger a_{q_2}^\dagger \langle m_1 b_1|g_{12}|m_2 p_1\rangle \langle p_1 q_1|g_{34}|b_1 q_2\rangle}{2\left(\varepsilon_{b_1}-\varepsilon_{p_1}-\varepsilon_{q_1}+\varepsilon_{q_2}\right)}$$

$$\frac{a_{m_1} a_{q_1} a_{n_2}^\dagger a_{q_2}^\dagger \langle m_1 b_1|g_{12}|p_1 n_2\rangle \langle p_1 q_1|g_{34}|b_1 q_2\rangle}{2\left(\varepsilon_{b_1}-\varepsilon_{p_1}-\varepsilon_{q_1}+\varepsilon_{q_2}\right)}$$

$$\frac{a_{m_1} a_{p_1} a_{q_1} a_{m_2}^\dagger a_{n_2}^\dagger a_{q_2}^\dagger \langle m_1 b_1|g_{12}|m_2 n_2\rangle \langle p_1 q_1|g_{34}|b_1 q_2\rangle}{2\left(\varepsilon_{b_1}-\varepsilon_{p_1}-\varepsilon_{q_1}+\varepsilon_{q_2}\right)}$$

$$-\frac{a_{n_1} a_{m_2}^\dagger \langle t_1|v_2|a_1\rangle \langle a_1 n_1|g_{12}|m_2 t_1\rangle}{2\left(\varepsilon_{a_1}-\varepsilon_{t_1}\right)}$$

$$\frac{a_{n_1} a_{n_2}^\dagger \langle t_1|v_2|a_1\rangle \langle a_1 n_1|g_{12}|t_1 n_2\rangle}{2\left(\varepsilon_{a_1}-\varepsilon_{t_1}\right)}$$

$$\frac{a_{m_1} a_{m_2}^\dagger \langle t_1|v_2|b_1\rangle \langle m_1 b_1|g_{12}|m_2 t_1\rangle}{2\left(\varepsilon_{b_1}-\varepsilon_{t_1}\right)}$$

$$-\frac{a_{m_1} a_{n_2}^\dagger \langle t_1|v_2|b_1\rangle \langle m_1 b_1|g_{12}|t_1 n_2\rangle}{2\left(\varepsilon_{b_1}-\varepsilon_{t_1}\right)}$$

$$\frac{a_{m_1} a_{p_2}^\dagger \langle t_1|v_2|p_2\rangle \langle m_1|v_1|t_1\rangle}{\varepsilon_{p_2}-\varepsilon_{t_1}}$$

$$-\frac{a_{m_1} a_{n_1} a_{m_2}^\dagger a_{p_2}^\dagger \langle t_1|v_2|p_2\rangle \langle m_1 n_1|g_{12}|m_2 t_1\rangle}{2\left(\varepsilon_{p_2}-\varepsilon_{t_1}\right)}$$

$$\frac{a_{m_1} a_{n_1} a_{n_2}^\dagger a_{p_2}^\dagger \langle t_1|v_2|p_2\rangle \langle m_1 n_1|g_{12}|t_1 n_2\rangle}{2\left(\varepsilon_{p_2}-\varepsilon_{t_1}\right)}$$

$$-\frac{a_{q_1} a_{m_2}^\dagger \langle a_1 b_1|g_{12}|m_2 t_1\rangle \langle t_1 q_1|g_{34}|a_1 b_1\rangle}{2\left(\varepsilon_{a_1}+\varepsilon_{b_1}-\varepsilon_{q_1}-\varepsilon_{t_1}\right)}$$

$$\frac{a_{q_1} a_{m_2}^\dagger \langle a_1 b_1|g_{12}|m_2 t_1\rangle \langle t_1 q_1|g_{34}|b_1 a_1\rangle}{2\left(\varepsilon_{a_1}+\varepsilon_{b_1}-\varepsilon_{q_1}-\varepsilon_{t_1}\right)}$$

$$\frac{a_{q_1} a_{n_2}^\dagger \langle a_1 b_1 | g_{12} | t_1 n_2 \rangle \langle t_1 q_1 | g_{34} | a_1 b_1 \rangle}{2 \left(\varepsilon_{a_1} + \varepsilon_{b_1} - \varepsilon_{q_1} - \varepsilon_{t_1}\right)}$$

$$-\frac{a_{q_1} a_{n_2}^\dagger \langle a_1 b_1 | g_{12} | t_1 n_2 \rangle \langle t_1 q_1 | g_{34} | b_1 a_1 \rangle}{2 \left(\varepsilon_{a_1} + \varepsilon_{b_1} - \varepsilon_{q_1} - \varepsilon_{t_1}\right)}$$

$$-\frac{a_{q_1} a_{p_2}^\dagger \langle a_1 | v_1 | t_1 \rangle \langle t_1 q_1 | g_{34} | p_2 a_1 \rangle}{\varepsilon_{a_1} + \varepsilon_{p_2} - \varepsilon_{q_1} - \varepsilon_{t_1}}$$

$$-\frac{a_{n_1} a_{p_2}^\dagger \langle a_1 n_1 | g_{12} | q_1 t_1 \rangle \langle t_1 q_1 | g_{34} | p_2 a_1 \rangle}{2 \left(\varepsilon_{a_1} + \varepsilon_{p_2} - \varepsilon_{q_1} - \varepsilon_{t_1}\right)}$$

$$\frac{a_{n_1} a_{p_2}^\dagger \langle a_1 n_1 | g_{12} | t_1 q_1 \rangle \langle t_1 q_1 | g_{34} | p_2 a_1 \rangle}{2 \left(\varepsilon_{a_1} + \varepsilon_{p_2} - \varepsilon_{q_1} - \varepsilon_{t_1}\right)}$$

$$-\frac{a_{n_1} a_{q_1} a_{m_2}^\dagger a_{p_2}^\dagger \langle a_1 n_1 | g_{12} | m_2 t_1 \rangle \langle t_1 q_1 | g_{34} | p_2 a_1 \rangle}{2 \left(\varepsilon_{a_1} + \varepsilon_{p_2} - \varepsilon_{q_1} - \varepsilon_{t_1}\right)}$$

$$\frac{a_{n_1} a_{q_1} a_{n_2}^\dagger a_{p_2}^\dagger \langle a_1 n_1 | g_{12} | t_1 n_2 \rangle \langle t_1 q_1 | g_{34} | p_2 a_1 \rangle}{2 \left(\varepsilon_{a_1} + \varepsilon_{p_2} - \varepsilon_{q_1} - \varepsilon_{t_1}\right)}$$

$$\frac{a_{m_1} a_{p_2}^\dagger \langle t_1 q_1 | g_{34} | p_2 b_1 \rangle \langle m_1 b_1 | g_{12} | q_1 t_1 \rangle}{2 \left(\varepsilon_{b_1} + \varepsilon_{p_2} - \varepsilon_{q_1} - \varepsilon_{t_1}\right)}$$

$$-\frac{a_{m_1} a_{p_2}^\dagger \langle t_1 q_1 | g_{34} | p_2 b_1 \rangle \langle m_1 b_1 | g_{12} | t_1 q_1 \rangle}{2 \left(\varepsilon_{b_1} + \varepsilon_{p_2} - \varepsilon_{q_1} - \varepsilon_{t_1}\right)}$$

$$\frac{a_{m_1} a_{q_1} a_{m_2}^\dagger a_{p_2}^\dagger \langle t_1 q_1 | g_{34} | p_2 b_1 \rangle \langle m_1 b_1 | g_{12} | m_2 t_1 \rangle}{2 \left(\varepsilon_{b_1} + \varepsilon_{p_2} - \varepsilon_{q_1} - \varepsilon_{t_1}\right)}$$

$$-\frac{a_{m_1} a_{q_1} a_{n_2}^\dagger a_{p_2}^\dagger \langle t_1 q_1 | g_{34} | p_2 b_1 \rangle \langle m_1 b_1 | g_{12} | t_1 n_2 \rangle}{2 \left(\varepsilon_{b_1} + \varepsilon_{p_2} - \varepsilon_{q_1} - \varepsilon_{t_1}\right)}$$

$$\frac{a_{q_1} a_{q_2}^\dagger \langle a_1 | v_1 | t_1 \rangle \langle t_1 q_1 | g_{34} | a_1 q_2 \rangle}{\varepsilon_{a_1} - \varepsilon_{q_1} + \varepsilon_{q_2} - \varepsilon_{t_1}}$$

$$\frac{a_{n_1} a_{q_2}^\dagger \langle a_1 n_1 | g_{12} | q_1 t_1 \rangle \langle t_1 q_1 | g_{34} | a_1 q_2 \rangle}{2 \left(\varepsilon_{a_1} - \varepsilon_{q_1} + \varepsilon_{q_2} - \varepsilon_{t_1}\right)}$$

$$-\frac{a_{n_1} a_{q_2}^\dagger \langle a_1 n_1 | g_{12} | t_1 q_1 \rangle \langle t_1 q_1 | g_{34} | a_1 q_2 \rangle}{2 \left(\varepsilon_{a_1} - \varepsilon_{q_1} + \varepsilon_{q_2} - \varepsilon_{t_1}\right)}$$

$$\frac{a_{n_1} a_{q_1} a_{m_2}^\dagger a_{q_2}^\dagger \langle a_1 n_1 | g_{12} | m_2 t_1 \rangle \langle t_1 q_1 | g_{34} | a_1 q_2 \rangle}{2 \left(\varepsilon_{a_1} - \varepsilon_{q_1} + \varepsilon_{q_2} - \varepsilon_{t_1}\right)}$$

$$-\frac{a_{n_1} a_{q_1} a_{n_2}^\dagger a_{q_2}^\dagger \langle a_1 n_1 | g_{12} | t_1 n_2 \rangle \langle t_1 q_1 | g_{34} | a_1 q_2 \rangle}{2 \left(\varepsilon_{a_1} - \varepsilon_{q_1} + \varepsilon_{q_2} - \varepsilon_{t_1}\right)}$$

$$-\frac{a_{m_1} a_{q_2}^\dagger \langle t_1 q_1 | g_{34} | b_1 q_2 \rangle \langle m_1 b_1 | g_{12} | q_1 t_1 \rangle}{2 \left(\varepsilon_{b_1} - \varepsilon_{q_1} + \varepsilon_{q_2} - \varepsilon_{t_1}\right)}$$

$$\frac{a_{m_1} a_{q_2}^\dagger \langle t_1 q_1 | g_{34} | b_1 q_2 \rangle \langle m_1 b_1 | g_{12} | t_1 q_1 \rangle}{2 \left(\varepsilon_{b_1} - \varepsilon_{q_1} + \varepsilon_{q_2} - \varepsilon_{t_1}\right)}$$

$$-\frac{a_{m_1} a_{q_1} a_{m_2}^\dagger a_{q_2}^\dagger \langle t_1 q_1 | g_{34} | b_1 q_2 \rangle \langle m_1 b_1 | g_{12} | m_2 t_1 \rangle}{2 \left(\varepsilon_{b_1} - \varepsilon_{q_1} + \varepsilon_{q_2} - \varepsilon_{t_1}\right)}$$

$$\frac{a_{m_1} a_{q_1} a_{n_2}^\dagger a_{q_2}^\dagger \langle t_1 q_1 | g_{34} | b_1 q_2 \rangle \langle m_1 b_1 | g_{12} | t_1 n_2 \rangle}{2 \left(\varepsilon_{b_1} - \varepsilon_{q_1} + \varepsilon_{q_2} - \varepsilon_{t_1}\right)}$$

$$-\frac{a_{m_1} a_{q_1} a_{p_2}^\dagger a_{q_2}^\dagger \langle m_1 | v_1 | t_1 \rangle \langle t_1 q_1 | g_{34} | p_2 q_2 \rangle}{\varepsilon_{p_2} - \varepsilon_{q_1} + \varepsilon_{q_2} - \varepsilon_{t_1}}$$

$$-\frac{a_{m_1} a_{n_1} a_{q_2}^\dagger a_{p_2}^\dagger \langle t_1 q_1 | g_{34} | p_2 q_2 \rangle \langle m_1 n_1 | g_{12} | q_1 t_1 \rangle}{2 \left(\varepsilon_{p_2} - \varepsilon_{q_1} + \varepsilon_{q_2} - \varepsilon_{t_1}\right)}$$

$$\frac{a_{m_1} a_{n_1} a_{q_2}^\dagger a_{p_2}^\dagger \langle t_1 q_1 | g_{34} | p_2 q_2 \rangle \langle m_1 n_1 | g_{12} | t_1 q_1 \rangle}{2 \left(\varepsilon_{p_2} - \varepsilon_{q_1} + \varepsilon_{q_2} - \varepsilon_{t_1}\right)}$$

$$-\frac{a_{m_1} a_{n_1} a_{q_1} a_{m_2}^\dagger a_{p_2}^\dagger a_{q_2}^\dagger \langle t_1 q_1 | g_{34} | p_2 q_2 \rangle \langle m_1 n_1 | g_{12} | m_2 t_1 \rangle}{2 \left(\varepsilon_{p_2} - \varepsilon_{q_1} + \varepsilon_{q_2} - \varepsilon_{t_1}\right)}$$

$$\frac{a_{m_1} a_{n_1} a_{q_1} a_{n_2}^\dagger a_{p_2}^\dagger a_{q_2}^\dagger \langle t_1 q_1 | g_{34} | p_2 q_2 \rangle \langle m_1 n_1 | g_{12} | t_1 n_2 \rangle}{2 \left(\varepsilon_{p_2} - \varepsilon_{q_1} + \varepsilon_{q_2} - \varepsilon_{t_1}\right)}$$

$$-\frac{a_{n_1} a_{q_2}^\dagger \langle a_1 n_1 | g_{12} | t_1 u_1 \rangle \langle t_1 u_1 | g_{34} | a_1 q_2 \rangle}{2 \left(\varepsilon_{a_1} + \varepsilon_{q_2} - \varepsilon_{t_1} - \varepsilon_{u_1}\right)}$$

$$\frac{a_{n_1} a_{q_2}^\dagger \langle a_1 n_1 | g_{12} | u_1 t_1 \rangle \langle t_1 u_1 | g_{34} | a_1 q_2 \rangle}{2 \left(\varepsilon_{a_1} + \varepsilon_{q_2} - \varepsilon_{t_1} - \varepsilon_{u_1}\right)}$$

$$\frac{a_{m_1} a_{q_2}^\dagger \langle t_1 u_1 | g_{34} | b_1 q_2 \rangle \langle m_1 b_1 | g_{12} | t_1 u_1 \rangle}{2 \left(\varepsilon_{b_1} + \varepsilon_{q_2} - \varepsilon_{t_1} - \varepsilon_{u_1}\right)}$$

$$-\frac{a_{m_1} a_{q_2}^\dagger \langle t_1 u_1 | g_{34} | b_1 q_2 \rangle \langle m_1 b_1 | g_{12} | u_1 t_1 \rangle}{2 \left(\varepsilon_{b_1} + \varepsilon_{q_2} - \varepsilon_{t_1} - \varepsilon_{u_1}\right)}$$

$$\frac{a_{m_1} a_{n_1} a_{q_2}^\dagger a_{p_2}^\dagger \langle t_1 u_1 | g_{34} | p_2 q_2 \rangle \langle m_1 n_1 | g_{12} | t_1 u_1 \rangle}{4 \left(\varepsilon_{p_2} + \varepsilon_{q_2} - \varepsilon_{t_1} - \varepsilon_{u_1}\right)}$$

$$-\frac{a_{m_1} a_{n_1} a_{q_2}^\dagger a_{p_2}^\dagger \langle t_1 u_1 | g_{34} | p_2 q_2 \rangle \langle m_1 n_1 | g_{12} | u_1 t_1 \rangle}{4 \left(\varepsilon_{p_2} + \varepsilon_{q_2} - \varepsilon_{t_1} - \varepsilon_{u_1}\right)}$$

$$\frac{\langle a_1 | v_1 | p_1 \rangle \langle p_1 | v_2 | a_1 \rangle}{\varepsilon_{a_1} - \varepsilon_{p_1}}$$

$$\frac{\langle a_1 b_1 | g_{12} | p_1 q_1 \rangle \langle p_1 q_1 | g_{34} | a_1 b_1 \rangle}{4 \left(\varepsilon_{a_1} + \varepsilon_{b_1} - \varepsilon_{p_1} - \varepsilon_{q_1}\right)}$$

$$-\frac{\langle a_1 b_1 | g_{12} | p_1 q_1 \rangle \langle p_1 q_1 | g_{34} | b_1 a_1 \rangle}{4 \left(\varepsilon_{a_1} + \varepsilon_{b_1} - \varepsilon_{p_1} - \varepsilon_{q_1}\right)}$$

$$\frac{\langle a_1 b_1 | g_{12} | q_1 p_1 \rangle \langle p_1 q_1 | g_{34} | a_1 b_1 \rangle}{4 \left(\varepsilon_{a_1} + \varepsilon_{b_1} - \varepsilon_{p_1} - \varepsilon_{q_1}\right)}$$

$$\frac{\langle a_1 b_1 | g_{12} | q_1 p_1 \rangle \langle p_1 q_1 | g_{34} | b_1 a_1 \rangle}{4 \left(\varepsilon_{a_1} + \varepsilon_{b_1} - \varepsilon_{p_1} - \varepsilon_{q_1}\right)}$$

$$\frac{\langle a_1 | v_1 | t_1 \rangle \langle t_1 | v_2 | a_1 \rangle}{\varepsilon_{a_1} - \varepsilon_{t_1}}$$

$$-\frac{\langle a_1 b_1 | g_{12} | q_1 t_1 \rangle \langle t_1 q_1 | g_{34} | a_1 b_1 \rangle}{2 \left(\varepsilon_{a_1} + \varepsilon_{b_1} - \varepsilon_{q_1} - \varepsilon_{t_1}\right)}$$

$$\frac{\langle a_1 b_1 | g_{12} | q_1 t_1 \rangle \langle t_1 q_1 | g_{34} | b_1 a_1 \rangle}{2 \left(\varepsilon_{a_1} + \varepsilon_{b_1} - \varepsilon_{q_1} - \varepsilon_{t_1}\right)}$$

$$\frac{\langle a_1 b_1 | g_{12} | t_1 q_1 \rangle \langle t_1 q_1 | g_{34} | a_1 b_1 \rangle}{2 \left(\varepsilon_{a_1} + \varepsilon_{b_1} - \varepsilon_{q_1} - \varepsilon_{t_1}\right)}$$

$$-\frac{\langle a_1 b_1 | g_{12} | t_1 q_1 \rangle \langle t_1 q_1 | g_{34} | b_1 a_1 \rangle}{2 \left(\varepsilon_{a_1} + \varepsilon_{b_1} - \varepsilon_{q_1} - \varepsilon_{t_1}\right)}$$

$$\frac{\langle a_1 b_1 | g_{12} | t_1 u_1 \rangle \langle t_1 u_1 | g_{34} | a_1 b_1 \rangle}{4 \left(\varepsilon_{a_1} + \varepsilon_{b_1} - \varepsilon_{t_1} - \varepsilon_{u_1}\right)}$$

$$-\frac{\langle a_1 b_1 | g_{12} | t_1 u_1 \rangle \langle t_1 u_1 | g_{34} | b_1 a_1 \rangle}{4 \left(\varepsilon_{a_1} + \varepsilon_{b_1} - \varepsilon_{t_1} - \varepsilon_{u_1}\right)}$$

$$-\frac{\langle a_1 b_1 | g_{12} | u_1 t_1 \rangle \langle t_1 u_1 | g_{34} | a_1 b_1 \rangle}{4 \left(\varepsilon_{a_1} + \varepsilon_{b_1} - \varepsilon_{t_1} - \varepsilon_{u_1}\right)}$$

$$\frac{\langle a_1 b_1 | g_{12} | u_1 t_1 \rangle \langle t_1 u_1 | g_{34} | b_1 a_1 \rangle}{4 \left(\varepsilon_{a_1} + \varepsilon_{b_1} - \varepsilon_{t_1} - \varepsilon_{u_1}\right)}$$